\documentclass[12pt]{article}

\usepackage{scicite}
\usepackage{times}
\usepackage{graphicx}  
\usepackage{dcolumn}   
\usepackage{bm}        
\usepackage{color}
\usepackage{amsmath}
\usepackage{amsfonts}
\usepackage{amssymb}
\usepackage{siunitx}

\usepackage{times}
\usepackage{epsfig}
\usepackage{epstopdf}
\usepackage{subfigure}
\usepackage{multirow}

\newenvironment{sciabstract}{%
\begin{quote} \bf}
{\end{quote}}

\title{Four-wave Mixing of Topological Edge Plasmons in Graphene Metasurfaces}

\author
{Jian Wei You,$^{1}$ Zhihao Lan,$^{1}$ and Nicolae C. Panoiu$^{1\ast}$\\
\\
\normalsize{$^{1}$Department of Electronic and Electrical Engineering, University College London,}\\
\normalsize{Torrington Place, London WC1E 7JE, United Kingdom}\\
\\
\normalsize{$^\ast$To whom correspondence should be addressed; E-mail: n.panoiu@ucl.ac.uk.} }

\date{}

\begin{document}


\baselineskip24pt


\maketitle


\begin{sciabstract}
We study topologically-protected four-wave mixing (FWM) interactions in a plasmonic metasurface
consisting of a periodic array of nanoholes in a graphene sheet, which exhibits a wide topological
bandgap at terahertz frequencies upon the breaking of time-reversal symmetry by a static magnetic
field. We demonstrate that due to the significant nonlinearity enhancement and large lifetime of
graphene plasmons in specific configurations, a net gain of FWM interaction of plasmonic edge
states within the topological bandgap can be achieved with pump power of less than
\SI{10}{\nano\watt}. In particular, we find that the effective waveguide nonlinearity coefficient
is about $\bm{\gamma}\simeq\SI{1.1e13}{\per\watt\per\meter}$, i.e., more than ten orders of
magnitude larger than that of commonly used, highly nonlinear silicon photonic nanowires. These
findings could pave a new way for developing ultra-low-power-consumption, highly-integrated and
robust active photonic systems at deep-subwavelength scale for applications in quantum
communications and information processing.
\end{sciabstract}

\section*{Introduction}
In the past decade, topological photonics has emerged as a rapidly burgeoning field of exploration
of topological physics in the context of photonics. This area of research began with the
theoretical work by Haldane and Raghu \cite{hr08PRL,rh08PRA}, where they constructed an analogue of
quantum Hall edge states in photonic crystals based on magneto-optical media and observed
topological edge modes within the corresponding photonic bandgaps. Shortly afterwards, an
experimental realization and observation of such topological edge modes in a magneto-optical
photonic crystal was reported in the microwave regime \cite{wcjs09Nature}. Since then, there has
been increasing interest in implementing in photonics topological states of matter as well as
developing new ideas specific to topological photonics
\cite{ljs14NatPhot,ljs16NatPhy,ks17NatPhot,lgyz17PRL,opag18arXiv}.

In addition to new perspectives brought in fundamental science, topological photonics also offers a
broad array of potential applications for novel photonic devices, as its exotic features have
already prompted the reexamination of some traditional views on light manipulation and propagation.
For instance, reducing back-reflection is a major challenge in optical waveguides, and in this
context the unidirectional topological waveguide \cite{wcjm08PRL,ypwg13APL,mffm14PRL} is an ideal
light transport device in integrated photonics, as it could transmit light without backscattering
even in the presence of inherent structural disorder. Moreover, some new concepts of topological
photonics have also led to the development of novel photonic devices, such as optical isolator
\cite{sbm18APL,hcll10APL}, robust delay lines \cite{hmfm13NatPhot,cjnm16NatMat}, signal switches
\cite{zj11JOSAB}, non-reciprocal devices \cite{fllr10APL,fllg11APL,qws11OE}, and topological lasers
\cite{bnve17Sci,sggl17NatPhot}.

Most of the previous studies have focused on linear topological photonic systems; however,
topological physics could also play an important role in the nonlinear regime, leading to novel
collective phenomena and strongly-correlated states of light \cite{opag18arXiv}. For instance, a
topological source of quantum light has recently been realized in a nonlinear photonic system,
which paves a new way for the development of robust quantum photonic devices \cite{mgh18Nat}.

\section*{Results}
\subsection*{The system}

We study a topologically-protected nonlinear four-wave mixing process in a graphene plasmonic
system. Graphene distinguishes itself as an ideal platform to study nonlinear topological photonics
in several key aspects: First, graphene exhibits remarkably large nonlinearity over a broad
spectral range, from terahertz to visible light. In particular, it has been shown \cite{yb12PRL}
that graphene in a strong magnetic field has the largest third-order susceptibility of all known
materials. Second, some recent studies \cite{jcsf17PRL,pyxa17NatComm} showed that topologically
protected one-wave edge plasmons can be realized in graphene metasurfaces. In addition to the
typical plasmonic effects, such as local-field enhancement and field confinement, the local field
can be further confined to the edge of the graphene plasmonic system, leading to a dramatic
enhancement of the nonlinear optical response of the graphene system. Third, phase-matching is a
crucial condition in nonlinear frequency mixing processes. In contrast to the frequently used bulk
modes \cite{mgh18Nat}, where several modes with different wave vectors usually exist at a given
frequency, a topological edge mode has a unique wave vector at a fixed frequency, thus the
phase-matching condition is insensitive to the way one excites the system, as in this case only a
single mode could be excited at one frequency. These important features make graphene plasmonic
systems particularly appealing in the design of highly integrated nonlinear topological
nanophotonic devices.
\begin{figure*}[t!]\centering
\includegraphics[width=13 cm]{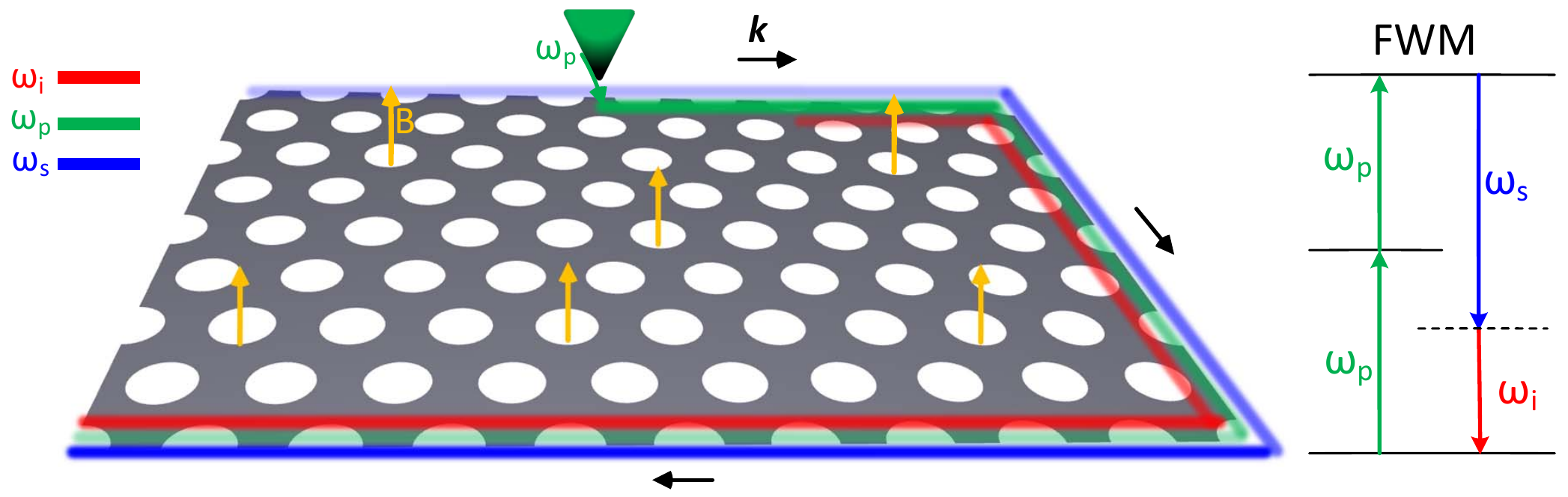}
\label{fig:PhysSche} \caption{Four-wave mixing of topologically protected one-way edge plasmons in
a graphene metasurface consisting of a periodic nanohole array with hexagonal symmetry in a static
magnetic field and the corresponding energy level diagram of the four-wave mixing process.}
\end{figure*}

The nonlinear system explored in this work is illustrated in Fig.~1. A graphene plasmonic
metasurface consisting of a periodic nanohole array with hexagonal symmetry is placed in a static
magnetic field. Due to the time-reversal-symmetry breaking induced by the magneto-optical response
of graphene under an external magnetic field, this plasmonic system could possess a topological
bandgap. After a geometry optimization, the topological bandgap could become wide enough so that
optical modes taking part in a four-wave mixing (FWM) process can readily fit in it. In order to
induce a FWM process, the system is excited by an external source at the pump frequency $\omega_p$,
as illustrated in Fig.~1. Due to the strong third-order nonlinearity of graphene, a degenerate FWM
process could take place, where two photons in a pump mode will generate a pair of photons at the
signal and idler frequencies, $\omega_s$ and $\omega_i$, respectively. As a result, the energy of
the pump wave (green) in Fig.~1 is transferred to the (seeded) signal (blue) and idler (red) modes,
leading to the pump decay and the amplification of the signal and idler. More importantly, this
degenerate FWM process is topologically protected by the chiral nature of the edge plasmons.

\subsection*{Topological bands of the graphene plasmonic system: linear response}

We first study the linear optical response of our graphene plasmonic system by calculating its
photonic band structure using a numerical approach based on the finite-element method (FEM). The
unit cell (with lattice constant $a$ and air hole radius $r$) and the first Brillouin zone of the
system used in our simulations are shown in Fig.~2(a) and 2(b), respectively. The band diagrams of
the system ($a=\SI{400}{\nano\meter}$, $r=\SI{120}{\nano\meter}$) at different magnetic field
$B=0,2,5,7,\SI{10}{\tesla}$ are presented in Fig.~2(c), where the parameters of the graphene are
set to be $E_F=\SI{0.2}{\electronvolt}$, $v_F\approx\num{}\SI{e6}{\meter\per\second}$ and
$\tau=\SI{50}{\pico\second}$ (see the section of Materials and Methods for details).
\begin{figure*}[b!]\centering
\includegraphics[width=16 cm]{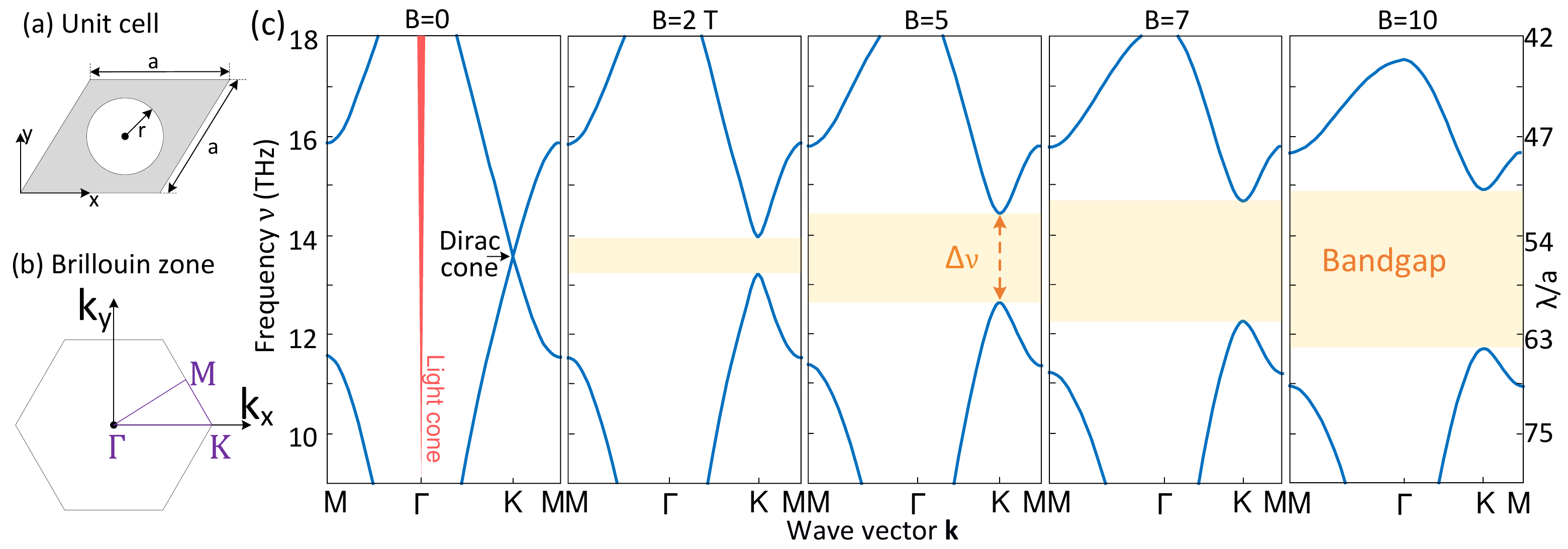}
\label{fig:BandInfSys} \caption{{\bf Band diagrams of the graphene plasmonic metasurface.} (a) Unit
cell and (b) the  first Brillouin zone of the metasurface. (c) Band diagrams of the metasurface at
$B=0, 2, 5, 7, \SI{10}{\tesla}$. As the Dirac cone is below the air light cone, surface plasmons
can exist around this cone at deep-subwavelength scale ($\lambda/a>40$). Moreover, a topological
bandgap is opened in the presence of an external static magnetic field.}
\end{figure*}

There are several notable features of the results presented in Fig.~2(c). First, without the
external magnetic field ($B=0$), due to the hexagonal symmetry of the metasurface structure, Dirac
cones that are protected by the parity ($P$) inversion and time-reversal ($T$) symmetries exist at
$K$ and $K^{\prime}$ symmetry points of the Brillouin zone \cite{jcsf17PRL}. Second, in the
presence of the magnetic field ($B\neq0$), the time-reversal symmetry of the system is broken and,
consequently, the Dirac cones are gapped out resulting in a topological nontrivial bandgap.
Moreover, the width $\Delta \nu$ of this bandgap increases as the amplitude of the magnetic field
increases.

To confirm the topological nature of the bandgap, we illustrate the emergence of the edge modes
within the bandgap for a finite size system along the $y$-axis, i.e., the number of unit cells
along this direction is finite (chosen to be $20$ in our FEM simulations), whereas the system is
periodic along the $x$-axis [in the FEM simulations, periodic boundary conditions are imposed at
the left and right boundaries along the $x$-axis, see Fig.~3(a)]. The supercell for this finite
system is shown by a green rectangle in Fig.~3(a), whose width and length are $a$ and
$b=\sqrt{3}a$, respectively.

The projected band diagrams along $k_x$, determined for $B=0,2,5,7,\SI{10}{\tesla}$, are depicted
in Fig.~3(b). First, similar to what we observed in Fig.~2(c), a bandgap opens when one applies an
external static magnetic field ($B\neq0$), with the gap width $\Delta\nu$ increasing as the
amplitude of magnetic field increases. However, different from the band diagrams of an infinite
graphene metasurface shown in Fig.~2(c), in the band diagrams of Fig.~3(b) there are two additional
edge modes at the top (red) and bottom (blue) boundaries of the finite graphene system. These two
edge modes connect the bulk bands located above and below the bandgap, and cannot be moved out of
the bandgap into the bulk bands as long as the bandgap exists. In other words, they are robust and
defect-immune, guaranteed by the topological protection of the bandgap.
\begin{figure*}[b!]\centering
\includegraphics[width=16 cm]{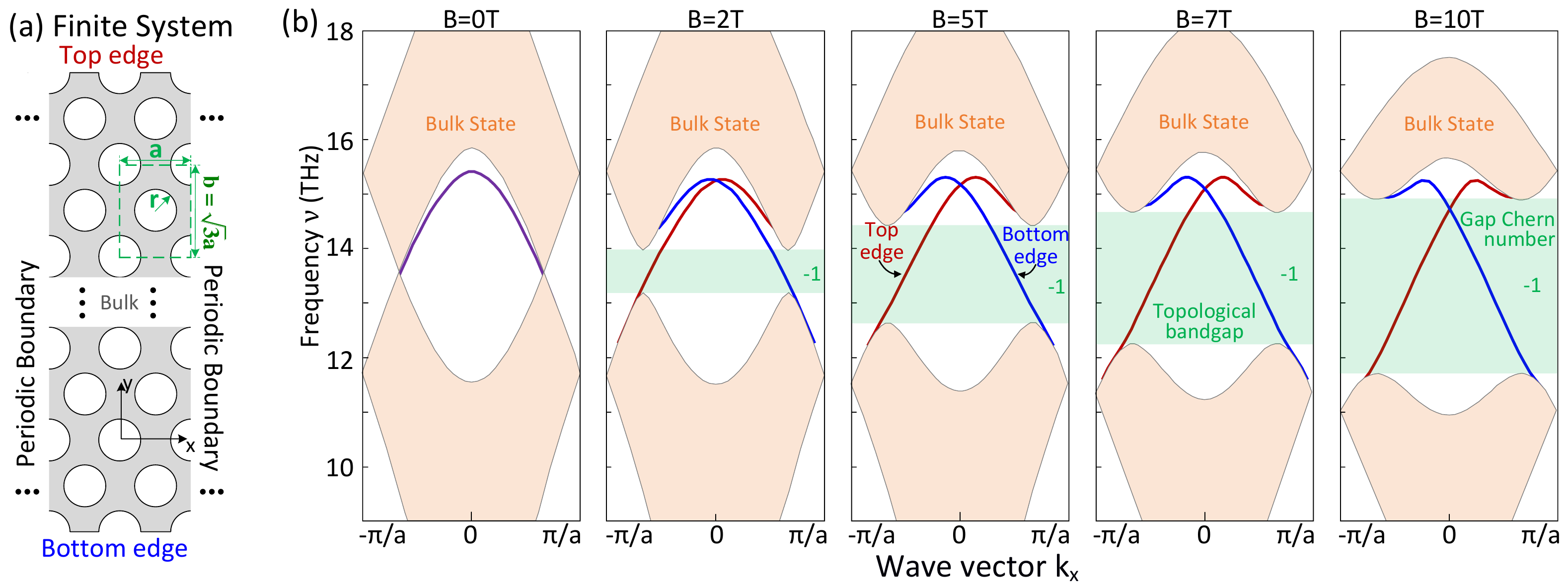}
\label{fig:BandFiniSys} \caption{{\bf Band diagrams of a finite graphene metasurface.} (a) Geometry
of the finite graphene metasurface, where the number of unit cells (green dashed frame) is finite
along the $y$-axis and infinite along the $x$-axis. (b) Projected band diagrams of the metasurface
at $B=0,2,5,7,$ and \SI{10}{\tesla}, where the edge modes on the top and bottom boundaries are
depicted by red and blue curves, respectively. Since the gap Chern number characterizes the number
of edge modes in the gap, there is a single edge mode at each boundary.}
\end{figure*}

One can also calculate the gap Chern number, which is a topological invariant that characterizes
the topological properties of the bandgap, to further confirm that the bandgap discussed above is
topologically nontrivial and that these edge modes are topological modes \cite{jcsf17PRL}. To this
end, we present in Fig.~3(b) (green) the calculated gap Chern number. Since the gap Chern number is
$-1$ (the magnitude indicates the number of topological edge modes, whereas the sign shows the
direction of propagation), there is only one topologically protected edge mode for each edge
termination. In other words, our graphene structure supports modes that can exhibit unidirectional
and defect-immune propagation features along the top and bottom edges. Moreover, the property of
unidirectional propagation of the edge modes is also illustrated by the slope of the edge modes in
the bandgap, as their group velocity, $v_g=\partial\omega/\partial k$, within the topological
bandgap is either positive (top edge) or negative (bottom edge).
\begin{figure*}[b!]\centering
\includegraphics[width=16 cm]{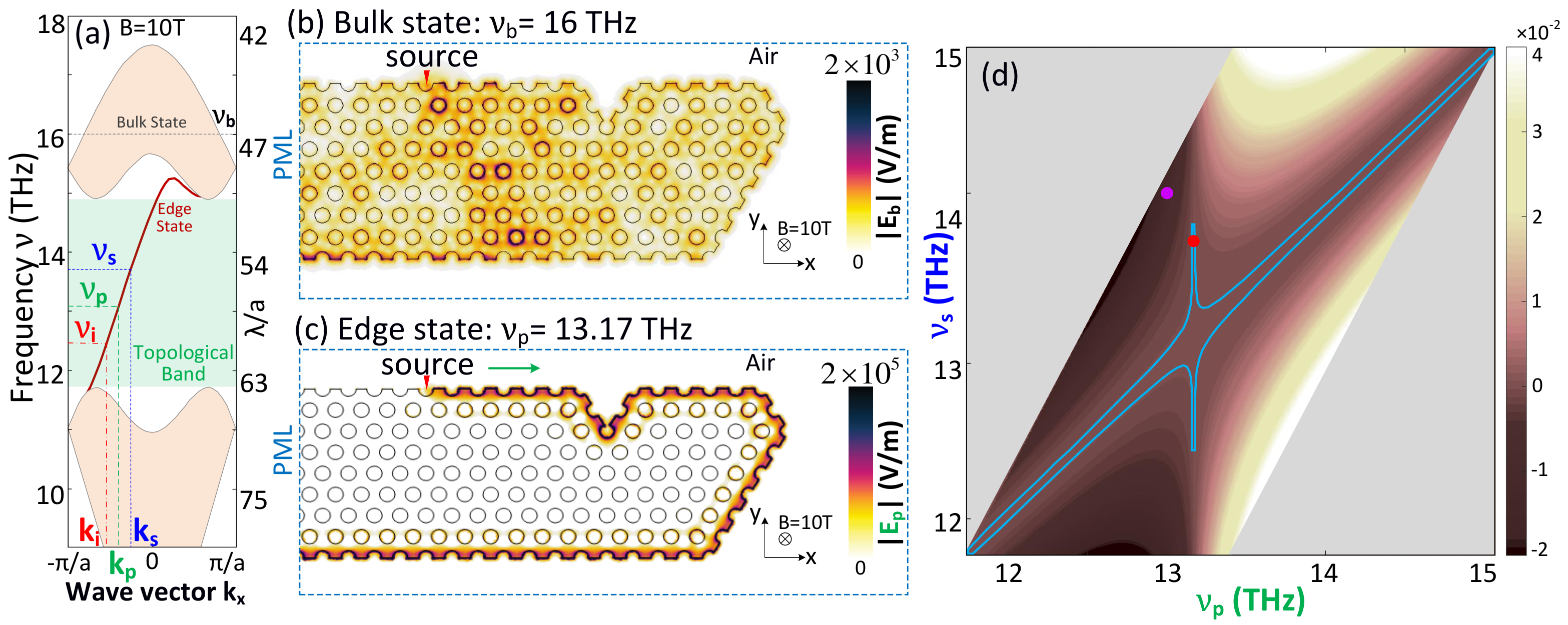}
\label{fig:FWMPhasMatch} \caption{{\bf Nonlinear edge-mode interaction and four-wave mixing within
a wide topological bandgap.} (a) Band diagram of the graphene metasurface at $B=\SI{10}{\tesla}$.
(b) Field profile of a bulk mode excitation at $\nu_b=\SI{16}{\tera\hertz}$, showing that the
optical field spreads throughout the bulk region. (c) Field profile of an edge mode excitation at
$\nu_b=\SI{13.17}{\tera\hertz}$ THz, illustrating its unidirectional and defect-immune propagation
along the system edge ($\tau\rightarrow\infty$). (d) Dispersion map of the normalized wave vector
mismatch $\Delta\kappa$. The blue contour is defined by the condition $\Delta\kappa=\num{e-5}$,
whereas the red and magenta dots correspond to a nearly phase-matched
($\Delta\kappa=\num{5.16e-6}$) and a phase-mismatched ($\Delta\kappa=\num{1.75e-2}$) FWM process,
respectively.}
\end{figure*}

In order to gain deeper physical insights into the physical properties of plasmonic bulk and edge
modes of the graphene metasurface, the near field distribution of these modes propagating in a
finite graphene plasmonic metasurface [4 unit cells along the $y$-axis and about $15$ unit cells
along the $x$-axis, as per Figs.~4(b) and 4(c)], is studied by using full-wave FEM simulations. In
the simulations, a perfectly-matched layer (PML) is used at the left side of the graphene
structure, whereas at the other sides we imposed scattering boundary conditions so as to mimic
infinite air space. In order to excite specific modes of this finite graphene system, an electric
source ($E_{0}=2\times\num{e4}\SI{}{\volt/\meter}$) depicted by a red triangle in Fig.~4(b) is
used.

In the case of the bulk mode, we choose the source frequency $\nu_b=\SI{16}{\tera\hertz}$, which
belongs to the bulk region [see Fig.~4(a)]. As expected, the corresponding optical field spreads
throughout the graphene structure [see Fig.~4(b)], which proves that indeed a bulk mode is excited
in this case. It should be noted that, due to the plasmonic characteristics of the graphene bulk
modes, their optical field is tightly confined at the surface of the graphene metasurface.

In the case of the excitation of an edge mode, we choose a frequency in the bandgap,
$\nu_p=\SI{13.17}{\tera\hertz}$, so that at this frequency only the edge mode exists, as per
Fig.~4(a). The corresponding field profile generated by the source in the finite system, presented
in Fig.~4(c), illustrates several notable features. Thus, the optical field does not penetrate in
the bulk region and only propagates unidirectionally along the edge of the graphene metasurface. In
addition, because of the chiral nature of the edge mode, this unidirectional propagation is robust
against structural defects, which allows it to circumvent defects (e.g., sharp bends) without
producing back-scattering. These features prove that the edge modes within the topological bandgap
in Fig.~4(a) are indeed topologically protected. Last but not least, we note that in addition to
the plasmonic field confinement effect illustrated in Figs.~4(b) and 4(c), the optical field of the
edge modes is further confined to the edge of the system, which is particularly important when one
seeks to enhance nonlinear optical interactions such as FWM.

\subsection*{Nonlinear interaction of edge states: Four-wave mixing}
Comparing Figs.~4(b) and 4(c), one can see that, in the case of the edge mode, the plasmon-induced
field-enhancement effect quantified by the ratio $|E_e|_{max}/E_{0}$, is two orders of magnitude
stronger than that in the case of bulk mode $|E_b|_{max}/E_{0}$, namely
$|E_e|_{max}/|E_b|_{max}>100$. Moreover, the results in Fig.~4(a) show that at a particular
frequency there is only a single edge mode, with a unique wave vector, whereas several different
bulk modes can be excited at one frequency. The former effect is important in enhancing the
nonlinear optical interactions, whereas the latter one is particularly useful for engineering
physical configurations in which phase-matching in FWM is achieved.

In order to illustrate these ideas, in what follows we analyze the circumstances in which the
phase-matching condition in a degenerate FWM of one-way edge modes can be fulfilled. To this end,
we calculate the normalized wave vector mismatch, defined as $\Delta\kappa=a\Delta
k=a(2k_{p}-k_{s}-k_{i})$, corresponding to a FWM process in which a pump edge mode with wave vector
$\mathbf{k}_{p}$ gives rise to signal and idler edge modes with wave vectors $\mathbf{k}_{s}$ and
$\mathbf{k}_{i}$, respectively. In particular, the exchange of energy among the interacting waves
is most efficient when the phase-matching condition $\Delta\kappa=0$ is satisfied. Unlike the wave
vector, the energy is conserved in the FWM process, meaning $2\nu_p=\nu_s+\nu_i$.

Starting from the mode dispersion curves of the topological edge modes presented in Fig.~4(a), and
using the energy conservation relation that characterizes the FWM process, the corresponding
dispersion map of the normalized wave vector mismatch $\Delta\kappa$ is calculated numerically and
depicted in Fig.~4(d). In particular, for the sake of a better quantitative understanding of the
energy conversion efficiency of the FWM process, we also show in this figure the contour defined by
$\Delta\kappa=\num{e-5}$. More specifically, for frequencies inside the domain defined by this
contour, energy is transferred from the pump to the signal and idler over a distance larger than
$\sim\num{e5}\pi$ lattice constants.

In order to validate the conclusions derived from the dispersion map of $\Delta\kappa$, we perform
full-wave simulations of the nonlinear dynamics of the interacting edge modes. To this end, we
chose a point indicated with a red dot in Fig.~4(d), characterized by
$\nu_p=\SI{13.17}{\tera\hertz}$, $\nu_s=\SI{13.72}{\tera\hertz}$, and
$\nu_i=\SI{12.62}{\tera\hertz}$, and for which the FWM interaction is nearly phase-matched
($\Delta\kappa=\num{5.16e-6}$). Moreover, we considered a seeded FWM process in which the input
intensity of the signal is much smaller than that of the pump, whereas the input intensity of the
idler is set to zero. Specifically, in our FEM simulations we set the source input field amplitudes
at the three frequencies $|E_p|=\SI{2e4}{\volt\per\meter}$, $|E_s|=\SI{4e2}{\volt\per\meter}$, and
$|E_i|=0$. Finally, the nonlinearity of graphene under the influence of a magnetic field of
\SI{10}{\tesla} is described by a third-order susceptibility with value of
$\chi^{(3)}=\num{}\SI{5e-10}{\square\meter\per\square\volt}$ \cite{yb12PRL}.

Using the procedure we just described, we have computed the near-field profiles at the frequencies
of the pump, signal, and idler, the results of these simulations being summarized in Fig.~5. In
these calculations graphene losses are neglected by setting $\tau\rightarrow\infty$. It can be seen
in Figs.~5(a) and 5(b) that, as the result of the nonlinear FWM interaction, the signal is
amplified upon propagation whereas an edge mode is generated at the idler frequency [note that in
Fig.~5(b) there is no external source at the idler frequency]. Equally important, since the
frequency of all the interacting edge modes are located in the topological bandgap, both signal and
idler modes are topologically protected and exhibit unidirectional and defect-immune propagation
along the system edge.
\begin{figure*}[t!]\centering
\includegraphics[width=15cm]{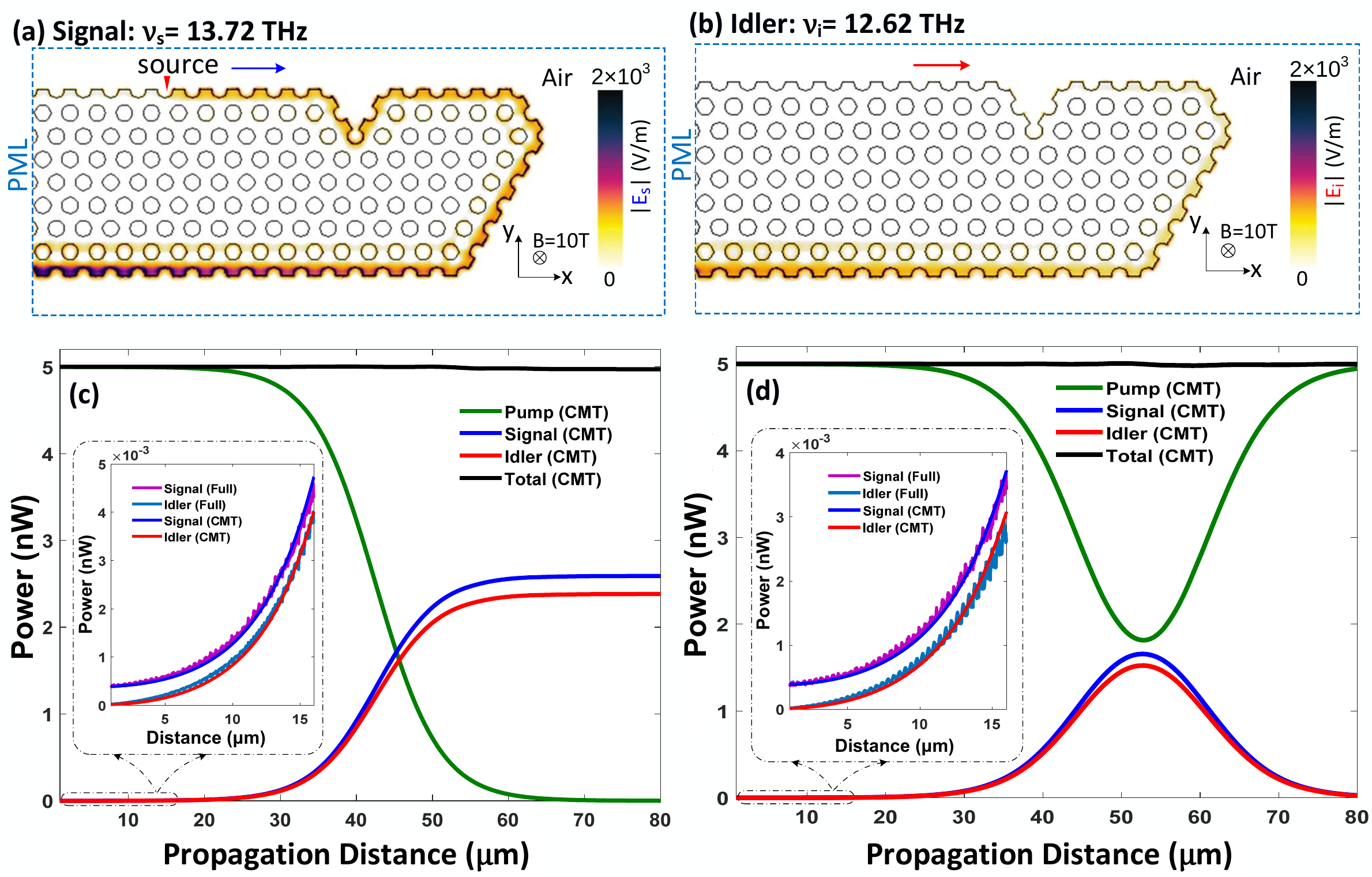}
\label{fig:EFieldDisFWM} \caption{{\bf Topologically-protected FWM process in a graphene
metasurface.} (a) The field profile at the signal frequency, $\nu_s=\SI{13.72}{\tera\hertz}$. (b)
The field profile at the idler frequency, $\nu_i=\SI{12.62}{\tera\hertz}$. (c) Dependence on the
propagation distance of the mode power of the pump, signal, and idler corresponding to the field
profiles shown in Fig.~5(a) and Fig.~5(b), determined using the CMT when the FWM process is
phase-matched. Also shown in the inset are the same mode powers determined using the CMT and
full-wave simulations. The black curve corresponds to the total power and shows that the energy is
conserved in the FWM interaction. (d) The same as in Fig.~5(c), but corresponding to a case when
the FWM interaction is not phase-matched.}
\end{figure*}

The FWM process can be further, quantitatively investigated by calculating the dependence on the
propagation distance of the power carried by the three edge modes. The mode power is calculated by
integrating the corresponding Poynting vector across the transverse section. The results of these
calculations are summarized in Figs.~5(c) and 5(d), and correspond to the case of near
phase-matching discussed above and a case when the FWM process is not phase-matched, respectively.
In the latter case, the system parameters are $\nu_p=\SI{13.03}{\tera\hertz}$,
$\nu_s=\SI{14.05}{\tera\hertz}$, $\nu_i=\SI{12.01}{\tera\hertz}$, and $\Delta\kappa=\num{1.75e-2}$
[the magenta point in Fig.~4(d)].

There are several important ideas revealed by the results presented in Figs.~5(c) and 5(d). First,
the power of both the signal and idler modes is amplified upon propagation, due to the energy
conversion from the pump mode. Second, the growth rate of signal and idler modes in the case of the
nearly phase-matched FWM is larger than that corresponding to the case when the FWM interaction is
not pase matched, which means that the energy conversion is more efficient in the former case.
Third, there is a very good agreement between the predictions of the coupled-mode theory (CMT) (see
Supplementary Materials for details) and the results obtained using full-wave simulations of the
nonlinear mode interaction, despite the fact that the optical fields at the three frequencies are
strongly confined at deep-subwavelength scale and significantly enhanced.

Importantly, our CMT predicts that the effective nonlinear FWM coefficient is
$\gamma_{FWM}\approx\SI{2.4e13}{\per\watt\per\meter}$ (see Supplementary Materials), which is more
than ten orders of magnitude larger than that of silicon photonic wire waveguides
\cite{ly16PRB,nsdm17Sci}. To the best of our knowledge, this is the largest nonlinear FWM
coefficient reported in a nonlinear optical system to date. This remarkable result is a consequence
of the particularly large third-order susceptibility of graphene, which is further enhanced by the
plasmon-induced enhancement and extreme confinement of the optical field of the edge modes. In
particular, the size of the unit cell of the graphene metasurface is much smaller than the
operating wavelength, namely $\lambda/a>50$ in our FWM process, a notable feature that can
facilitate the design of low-power, ultracompact active photonic nanodevices.

The radiation loss of the edge modes of the graphene metasurface can be neglected because, as we
just discussed, they are strongly confined. The intrinsic loss, however, has to be taken into
account in practical applications. To study its influence on the FWM interaction, a finite plasmon
lifetime $\tau$ is considered in Eq.~2 and Eq.~3. Similarly to the lossless case, we determined the
dependence of the power of the interacting edge modes on the propagation distance, the
corresponding results of this analysis being presented in Fig.~6.

Typically, the plasmon lifetime is determined by the plasmon-phonon coupling and varies from
\SIrange{0.1}{1}{\pico\second} \cite{la14ACSNano}. This loss can be reduced if exfoliated graphene
is placed onto a boron nitride substrate \cite{dyml10NatNano}, which leads to a lifetime as large
as $\SI{3}{\pico\second}$. Moreover, recent experiments \cite{yllz12Nanolett,pwbp13PRL} have
demonstrated that an external magnetic field can also strongly increase the plasmon lifetime, as in
this case the two-dimensional surface plasmon can be effectively transformed into a
one-dimensional-like edge plasmon. Specifically, it has been shown that by applying an external
magnetic field the plasmon lifetime can be readily increased to $\SI{50}{\pico\second}$
\cite{pwbp13PRL}.
\begin{figure*}[b!]\centering
\includegraphics[width=15 cm]{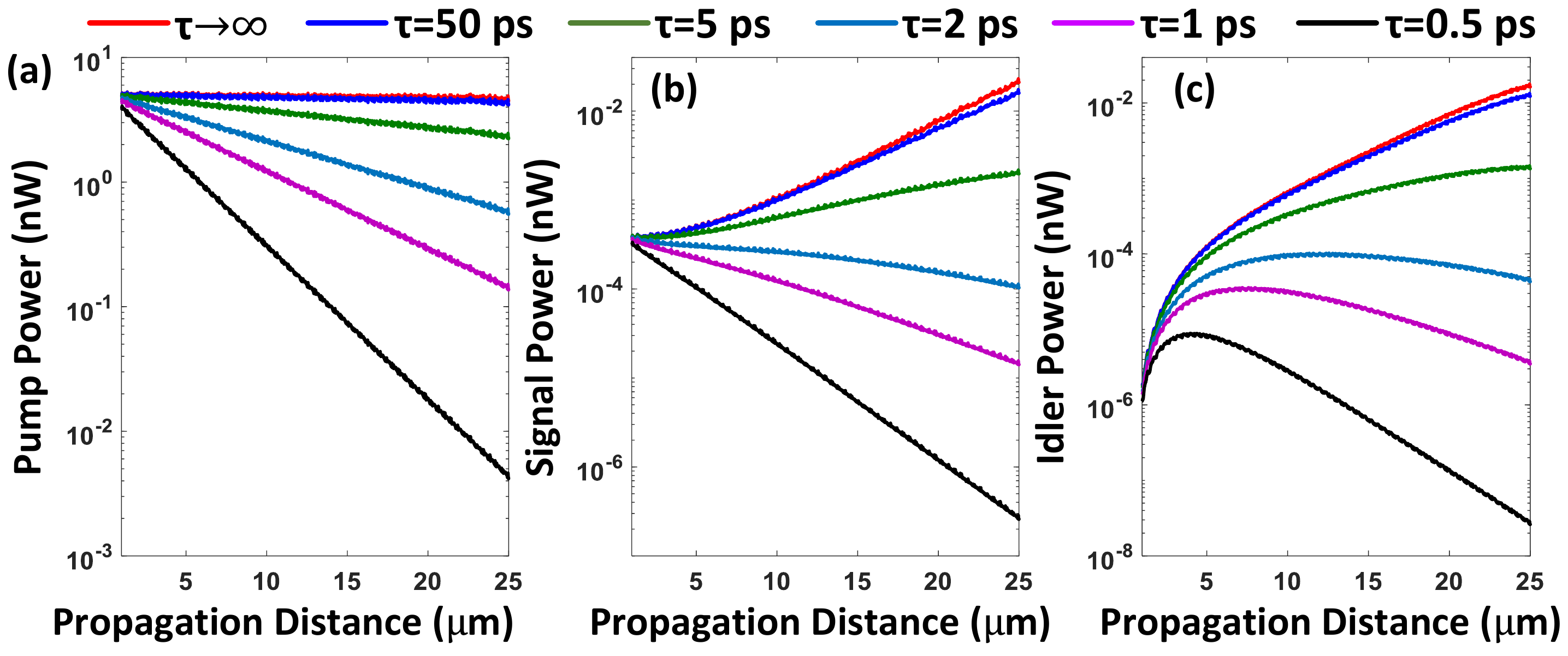}
\label{fig:EFieldDisFWM} \caption{{\bf Influence of loss on the topologically-protected FWM process
in graphene metasurface at $B=\SI{10}{\tesla}$.} (a), (b), (c) Dependence on the propagation
distance of the power of the pump, signal, and idler edge modes, respectively, corresponding to a
phase-matched FWM process determined for different values of the loss rate. Owing to the
particularly large value of the FWM coefficient, net gain (the FWM gain overcompensates the loss)
can be achieved as long as $\tau\gtrsim\SI{2.5}{\pico\second}$.}
\end{figure*}

One remarkable conclusion of the analysis of the FWM interaction of edge modes of lossy graphene is
that net signal gain can be achieved if the life-time $\tau\gtrsim\SI{2.5}{\pico\second}$. In fact,
this is the first plasmonic system in which net gain can be achieved without incorporating in the
system gain optical media. Indeed, it can be seen in Fig.~6 that whereas the pump decays for all
values of the plasmon life-time, due to the combined contributions of the graphene loss and energy
transfer mediated by the FWM interaction, the signal power increases monotonously if
$\tau\gtrsim\SI{2.5}{\pico\second}$. The idler, on the other hand, shows a more complex dynamics.
Thus, irrespective of the value of the life-time, at the beginning of the nonlinear interaction the
power in the idler builds up over a certain distance. After this amplification stage the power in
the idler decays monotonously if $\tau\lesssim\SI{2.5}{\pico\second}$, because the pump power is no
longer large enough to sustain the amplification of the idler, whereas if
$\tau\gtrsim\SI{2.5}{\pico\second}$ the power in the idler mode increases monotonously over the
entire distance considered in our simulations.

\subsection*{Conclusion}
Using rigorous full-wave simulations supported by a coupled-mode theory we have demonstrated a
topologically protected nonlinear four-wave mixing process in a patterned graphene plasmonic
metasurface. In particular, we have shown that a topological bandgap as wide as several THz can be
created in the metasurface under a strong static magnetic field. Moreover, the analysis of the
dispersion properties of the topologically-protected edge modes located in the bandgap reveals that
four-wave mixing interaction is efficiently phase-matched in a large domain of the parameter space
of the system. The near-field profiles of the interacting edge modes show unidirectional and
defect-immune propagation, hence demonstrating that the four-wave mixing process is topologically
protected. Remarkably, our study also reveals that, due to an unusually large value of the
four-wave mixing nonlinear coefficient and the large field enhancement at plasmon resonances, the
four-wave mixing interaction produces net gain even when plasmon losses in graphene are rigorously
taken into account. This striking property of four-wave mixing of topological edge modes of
graphene metasurfaces might play an important r\^{o}le in the development of new ultracompact and
topologically-protected active photonic systems.

\subsection*{Materials and Methods}
In our modeling of an infinite graphene metasurface, periodic boundary conditions are used for the
four edges of the unit cell depicted in Fig.~2(a). At infrared and terahertz frequencies, graphene
placed in a static magnetic field can be characterized as an electrically gyrotropic material
\cite{clwo11NatPhy,fpn12PRB,ybbp18Nanophot}, whose surface conductivity tensor can be represented
as:
\begin{equation}\label{eq:SigmaTensor2}
\bm{\sigma_{s}}  = \left[ {\begin{array}{*{20}{c}}
{{\sigma _{L}}}&{{\sigma _{H}}}\\
{{-\sigma _{H}}}&{{\sigma _{L}}}
\end{array}} \right],
\end{equation}
where the diagonal elements (longitudinal conductivity, $\sigma_{L}$) and the off-diagonal elements
(Hall conductivity, $\sigma_{H}$) can be determined using Kubo's formalism \cite{dg02book}. At room
temperature and for frequencies below the visible-light region, the longitudinal and Hall
conductivities are given by \cite{jcsf17PRL,pyxa17NatComm}:
\begin{equation}\label{eq:SigmaL}
\sigma_L=\sigma_0\frac{1-i\omega\tau}{{(\omega_c\tau)}^2-{(i+\omega\tau)}^2},
\end{equation}
\begin{equation}\label{eq:SigmaH}
{\sigma _H} = -\sigma_0\frac{\omega_c\tau}{{(\omega_c\tau)}^2-{(i+\omega\tau)}^2},
\end{equation}
where $\sigma_0=e^2E_F\tau/(\pi\hbar^2)$, $\tau$ is the relaxation time (plasmon lifetime),
$\omega_c \approx e B_\perp v_F^2/E_F$ is the cyclotron frequency, with $B_\perp$, $v_F$, and $E_F$
being the external static magnetic field perpendicular onto the graphene surface, the graphene
Fermi velocity, and the Fermi energy, respectively.

The surface conductivities in Eq.~2 and Eq.~3 show that the graphene under a static magnetic field
($B_{\perp}\neq0$) is anisotropic, lossy, and dispersive. As a consequence, most of the traditional
electromagnetic eigenmode solvers based on the plane wave expansion method cannot be used to
determine the photonic band structure. In order to circumvent this problem, we used the numerical
solver of Comsol based on the FEM method to calculate the band diagrams of the graphene
metasurfaces investigated in this study. In our full-wave simulations, electric sources are placed
at positions with low symmetry, so that all modes are excited. Moreover, multiple probe monitors
are placed at low-symmetry locations, too, to determine the mode frequencies.

In the nonlinear simulations, three nonlinear surface currents are defined in the Comsol software,
namely one for the pump frequency ($\nu_p$), one for the signal frequency ($\nu_s$), and one for
the idler frequency ($\nu_i$). These nonlinear currents are coupled, as described by the following
equations:
\begin{equation}\label{eq:EqsEp}
J_p^{surf} =  6{\sigma_{p,surf}^{(3)}}{E_s}{E_i}E_p^*,
\end{equation}
\begin{equation}\label{eq:EqsEs}
J_s^{surf} =  3{\sigma_{s,surf}^{(3)}}{E_p}{E_p}E_i^*,
\end{equation}
\begin{equation}\label{eq:EqsEi}
J_i^{surf} =  3{\sigma_{i,surf}^{(3)}}{E_p}{E_p}E_s^*.
\end{equation}
Here, the third-order surface conductivity is defined as
$\sigma_{\alpha,surf}^{(3)}=-i\epsilon_0\omega_\alpha h_{eff} \chi^{(3)}$, $\alpha=p,s,i$, where
$\chi^{(3)}$ is the third-order bulk susceptibility and the thickness of graphene is assumed to be
$h_{eff}=\SI{0.3}{\nano\meter}$ \cite{yb12PRL}. Moreover, the electric fields $E_\alpha$,
$\alpha=p,s,i$, are the amplitudes of the pump, signal, and idler, respectively. More details about
the nonlinearity of magnetized graphene and the coupled-mode theory describing the FWM interaction
of graphene edge modes can be found in the Supplementary Materials.


\section*{Acknowledgments}
\paragraph*{Funding:} This work was supported by European Research Council (ERC), Grant Agreement no. ERC-2014-CoG-648328.
\paragraph*{Contributions:} N.C.P. conceived the idea and supervised the project. J.W.Y. and Z.L.  performed the numerical simulations. All authors contributed to the preparation of the manuscript.
\paragraph*{Competing interests:} The authors declare that they have no competing interests.
\paragraph*{Data and materials availability:} All data needed to evaluate the conclusions in the paper are present in the paper and/or the Supplementary Materials. Additional data related to this paper may be requested from the authors.

\clearpage

\end{document}